# Tailored optical properties of atomic medium by a narrow bandwidth frequency comb

Rita Behera[1,2], Bappaditya Pal[2], Swarupananda Pradhan[1,2,3]

[1]Homi Bhabha National Institute, Department of Atomic Energy, Mumbai-400094, India

[2]Beam Technology Development Group, Bhabha Atomic Research Centre, Mumbai-400085, India

[3]Correspondence address: spradhan@barc.gov.in and pradhans75@gmail.com

**Abstract**

The quantum interference assisted enhanced optical activity due to the emergence of a steady-state atomic polarization is investigated. The Rubidium atoms in an antirelaxation coated cell provide a suitable platform to address the phenomena at multiple Larmor's frequencies. It interacts with a narrow bandwidth frequency comb generated by the frequency modulation of the light field. The Lindblad master equation with a trichromatic field provides a microscopic picture of the atomic response to the narrow bandwidth frequency comb. The directive of the relative phase between the light fields, in the detuning dependence of the magnetic resonances, is conclusively captured with the trichromatic field model. The measured absorption, nonlinear magneto-optic rotation, and their dependencies on various experimental parameters are analysed. The ellipticity of the light field controls the extent of several physical processes at multiple Larmor's frequencies. The investigation provides an approach to address the Zeeman coherence in the interaction of a narrow bandwidth frequency comb with an atomic ensemble and will have applications in various quantum devices.

## Introduction:

The change in the polarization state of the light field is a sensitive means of extracting the material properties and has a crucial role in high sensitive magnetometry to cosmology [1-4]. It has been engineered by controlling the optical activity in the vicinity of atomic resonances with the parameters of the input light field and impinging magnetic field. The light amplitude and its polarization gets further altered over a narrow span of frequency around a two-photon Raman resonance [3-5]. It improves the sensitivity, precision, and flexibility at the application end. The emergence of a long-lived Zeeman coherence is crucial for the sub-luminal propagation of light, enhanced optical activity, superposition of states, qubits and optical computation, and others [3-9]. A further fascinating area of research can be the exploration of the Zeeman coherence in the interaction of a frequency comb with an atomic ensemble. The frequency comb spectroscopy has drawn prolonged interest due to its scope in advanced technologies [10,11].

An optical frequency comb consists of a series of coherent, evenly spaced discrete spectral lines. It conventionally refers to the frequency spectrum generated by a mode-locked pulse laser system. Another convenient way to produce the frequency comb is by modulating the laser frequency using external modulators like an electro-optic modulator [12]. The current modulation of a diode laser (avoids the external modulator) also produces a series of equally spaced light field in the frequency domain. The coherence is preserved over the whole span of the spectrum. It has a smaller bandwidth compared to the spectrum of a mode-locked pulsed laser system. So, the frequency modulation (FM) of a continuous-wave laser produces a narrow bandwidth frequency comb (NBFC). Moreover, in this method, the control over teeth separation is very simple because there is no need to handle sophisticated laser cavity. Since the frequency separation between the teeth component can be made very small, it provides a way to establish Zeeman coherence at small values of Larmor's frequency ($\Omega_L$). The NBFC can have a complementary role to the conventional optical frequency comb generated by the mode-locked laser system.

The equal spacing between the coherent teeth components of NBFC makes it a natural choice for the study of Raman resonance. The quantum interference leads to the establishment of Zeeman coherence as the atomic system interacts with multiple light fields. The use of a frequency-modulated light field induces temporal oscillations in the atomic polarization. These oscillations are synchronous with the applied modulation. They get resonantly enhanced as $\Omega_L$ approaches a multiple of the modulation frequency [13-18]. The erstwhile investigations have focussed on the oscillating atomic polarization induced due to the modulation in the light field. To our best knowledge, the microscopic act of the teeth components (of the comb) in the process is yet to be explored.

Recently, the emergence of a steady-state atomic polarization at a series of $\Omega_L$ separated by a half-integral multiple of the modulation frequency ($\omega_m$) has been reported [19]. The utilized frequency modulation produces a NBFC with a span much larger than the associated Doppler width. The separation between the individual frequency teeth is much smaller than the natural line width. Thus, the central (tail) part of the frequency comb is resonant with the atomic ensemble for an on (off) resonant light field. The distinct phase relationship between the neighbouring teeth is presumed to be responsible for the odd magnetic resonances (from the central component). The interaction of a very large number of light fields with the atomic system is a complicated process. A series of coupled closed-loop excitation participate in the process, and it is difficult to get a microscopic picture of the phenomena.

The objective of the present work is to study the nonlinear magneto-optic rotation (NMOR) of an atomic medium interacting with a NBFC. The complexity of the process is approximated to the interaction of a trichromatic field with the atomic ensemble. The Lindblad master equation with the trichromatic field provides the desired microscopic picture of the phenomena. The calculated profile consistently demonstrates the important role of the phase relationship between the teeth components. The calculated and measured profiles resemble well for the utilized modulation parameters in this work. The changes in absorption and NMOR by the atomic ensemble are measured for different ellipticities of the NBFC. The calculated line shapes provide a reference to analyse the role of optical pumping, absorption, dichroism, and birefringence behind the observed enhanced optical activity at multiple $\Omega_L$. Interestingly, the relative contribution of these processes strongly depends on $\Omega_L$, even under the same experimental condition.

## Experimental apparatus:

The experiment uses a vertical-cavity surface-emitting diode laser (VCSEL) tuned to the Rubidium (Rb) D1 atomic transition with an intensity of ~700 µW/cm$^2$. The laser is frequency modulated at $\omega_m$=12 kHz and modulation amplitude ($A_m$) ~ 1.63 GHz to generate the NBFC. It also facilitates laser frequency stabilization (as shown in the shaded part of Fig. 1) [18-20]. The laser frequency is locked at $\Delta$~+140 MHz from the $^{85}$Rb $F = 3 \rightarrow F' = 2$ transition, where $\Delta$ is the detuning of the central component of the NBFC from the atomic resonance. An anti-relaxation coated atomic cell at 24$^0$C is kept in a three-axis magnetic field-controlled environment. The measured Doppler width ($\Delta\gamma_D$) is ~450 MHz. The orthogonal magnetic fields are compensated by two pairs of rectangular coils along x and y directions, and four layers of mu-metal sheet enclosing the system. The NBFC, after interaction with the atomic ensemble, is analysed through a polarization beam splitter cube (PBS) marked as PBS-d. The axis of the PBS-d is same as the input PBS (PBS-i). The ellipticity of the light field is controlled by rotating the quarter-wave plate (QWP) by an angle (θ) with respect to the axis of PBS-i. A sinusoidal magnetic field @ 55Hz with an amplitude ~200 nT is added to the Bz (scanning) magnetic field. The transmitted and reflected light across the PBS-d, are phase sensitively detected in reference with the magnetic field modulation and are termed as MMzT and MMzR signal respectively. The magnetic coils are spectroscopically calibrated by using the splitting between the coherent population trapping (CPT) resonances [21,22]. It involves the application of a fixed current through each set of the orthogonal coils, measurement of the corresponding frequency spacing between the CPT resonances, and final calibration using atomic parameters of Rb atoms [23].

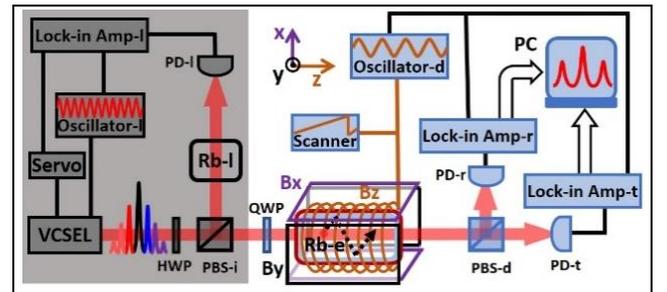

**Fig. 1:** Schematic diagram of the experimental set-up for study of optical activity near magnetic resonance using NBFC. The shaded part of the set-up is used for laser frequency stabilization. The amplitude and polarization of the light field after interaction with the atomic sample is phase sensitively detected by a polarimetric set-up. The extension i, l, d, r, and t are made to indicate input, locking, detection, reflected and transmitted respectively.

## Theoretical model:

The objective of this section is to develop a simple model to study the tailored optical activity by the NBFC. The overall process is a convolution of both dichroism and birefringence. The attributes of these processes are different for the on-resonant and off-resonant light field. Since the individual teeth components are narrowly separated in the frequency domain, both the resonant and off-resonant behaviour will contribute to the signal simultaneously. The optical pumping induced population imbalance between the Zeeman sub-states plays a prominent role for an elliptically polarized light. In the present circumstances, several competing close-loop excitations are in action in a Doppler broadened system. The phases of the individual teeth components add further complexity to the problem. Thus, getting a complete physical picture of the underlying mechanism is a tedious task. We found that a model with a trichromatic field can suitably explain the experimental observation. It provides a microscopic picture of the associated physical processes. The simplicity of the model demonstrates the generality of the observed phenomena.

The frequency spectrum of the NBFC depends on the modulation parameters ($\omega_m$ and $A_m$). The NBFC follows the properties of the Bessel's function of the first kind ($J_n(I_m)$), where $n$ and $I_m$ represent its order and modulation index respectively. The amplitude of the $n$th teeth is $\propto |J_{-n}(I_m)|^2$ and the phase relationship between the teeth is given by $J_{-n}(I_m) = (-1)^n J_n(I_m)$. The separation between the neighbouring teeth components is equal to $\omega_m$ and the span of the NBFC is proportional $A_m$. In our experiments, the values of $\omega_m$ and $A_m$ are 12 kHz and ~1.63 GHz respectively. So, a large number of frequency teeth are present in the Doppler profile ($\Delta\gamma_D$~450 MHz) of the atomic ensemble. However, compared to the total span of the NBFC (~ GHz), the Doppler envelope (~MHz) encompasses only a fraction of its spectrum (Fig.2A). As a result, the atomic sample is in resonance with only a specific spectral part of the NBFC in the frequency domain. As we vary $\Delta$, a different frequency region of the NBFC is in resonance with the atomic sample. Here, $\Delta$ is the frequency difference between the central component of the NBFC and the atomic resonance.

Fig. 2A indicates the variation in the relative phase and amplitude between the teeth components of the NBFC in different detuning regions, namely $\Delta=0$, $\Delta <0$, and $\Delta >0$. In the latter two conditions, the wings of NBFC are in resonance with the atomic sample. The central component of the NBFC is shown in black colour. Also, this component is pointed 'up' about the central horizontal grey line. The red and blue lines indicate 'odd' and 'even' teeth components respectively about the central component. These components are directed either up or down about the horizontal grey line. When a component is 'up' about the horizontal grey line, it has the same phase as the central component, which we have assigned as the '+ ve' phase. Similarly, when a teeth component is 'down', it has the opposite phase with the central component. We have assigned this as the '- ve' phase.

In the $\Delta=0$ region, for any three consecutive frequency components, the first and third components are always in opposite phase, irrespective of the second or middle component. It arises from the property of the Bessel's function as illustrated in Fig. 2A. Following our sign convention for phases as mentioned in the earlier paragraph, the phases of any three consecutive frequency component can only have one of the four forms (-,+,+), (+,+,-), (-,-,+) and (+,-,-). In contrast, for both $\Delta >0$ and $\Delta <0$ region, for any three consecutive frequency components, the first and third components always have the same phase. Consequently, the phases are in the forms (+,+,+), (-,-,-), (-,+,-), and (+,-,+).

These two classes of phase relationships display contrasting quantum interference at $\Omega_L = \pm \omega_m/2$.

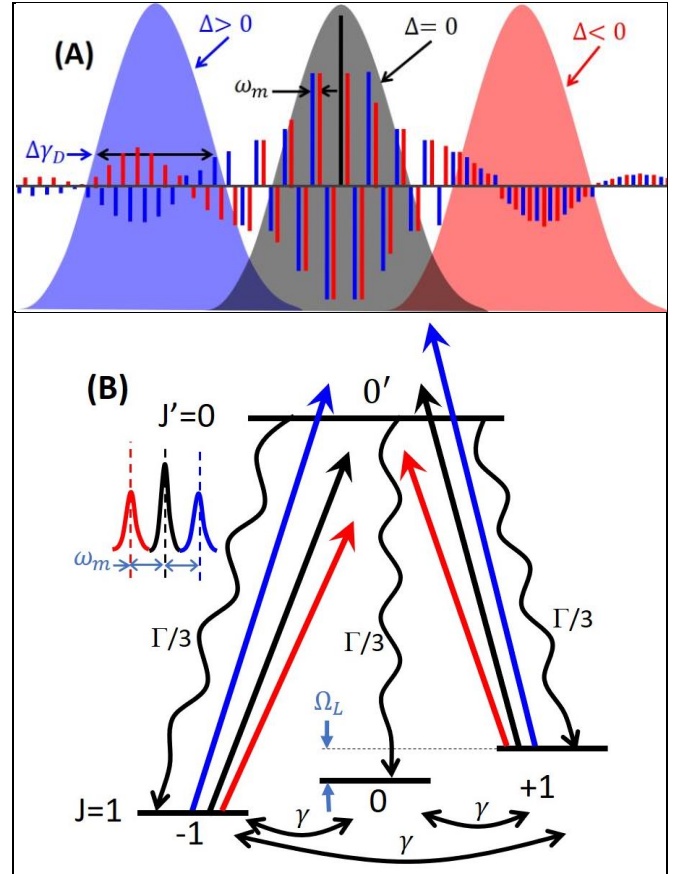

**Fig. 2:** A: The position of the Doppler profile in the NBFC for different detuning ($\Delta$). The phase relationship between the teeth changes from the centre to the wing of the comb and has important consequences. The black, red, and blue lines represent central, odd, and even teeth respectively. The height of the line indicates the amplitude of a teeth and up/down direction from the baseline represents its phase. The diagram is not to the scale and is illustrative only. The experimental value of Doppler width ($\Delta\gamma_D$) ~450 MHz and separation between teeth ($\omega_m$)=12kHz. B: The coupling diagram for $J = 1 \rightarrow J' = 0$ with a trichromatic field separated by $\omega_m$. The quantum interference between different sets of light fields is realized at multiple $\Omega_L$.

A theoretical model with J=1 to J=0 is appropriate for analysis of atomic response to the NBFC. The two-photon Raman resonance can be established at $\Omega_L = 0, \pm \omega_m/2$, and $\omega_m$ as can be realized from the schematic diagram in Fig. 2B. The model is developed based on Lindblad master equation $\frac{d\rho}{dt} = -\frac{i}{\hbar}[H,\rho] + \mathcal{L}(\rho)$. The algorithm is a standard practice in past literature and has been explicitly discussed for a variety of experimental conditions [24-30]. Here, we provide a brief summary of the procedure. The total Hamiltonian includes the bare Hamiltonian, laser atom interaction term, and the magnetic field interaction term ($H = H_0 + H_{int} + H_B$). The decay matrix $\mathcal{L}(\rho)$ takes care of spontaneous decay ($\Gamma$) and ground state collisional redistribution($\gamma$) associated with the relevant density matrix element ($\rho$). The optical field in spherical coordinates with an ellipticity $\epsilon$ is represented as

$$E = -(cos\,\epsilon + sin\,\epsilon)E_+\hat{\sigma}_+e^{-i\omega t} + (cos\,\epsilon - sin\,\epsilon)E_-\hat{\sigma}_-e^{-i\omega t} + c.c \quad (1)$$

where $E_+$ and $E_-$ are the electric field associated with the $\sigma_+$ and $\sigma_-$ light coupled to $m_J = -1 \rightarrow m_{J'} = 0$ and $m_J = 1 \rightarrow m_{J'} = 0$ respectively. For a trichromatic field with the $\sigma_+$ and $\sigma_-$ components of equal amplitudes ($E_+ = E_- = \mathcal{E}$), the Eq.1 becomes

$$E = \mathcal{E}(-(cos\,\epsilon + sin\,\epsilon)\hat{\sigma}_+ (e^{i\varphi_1} + e^{i\varphi_2}e^{i\omega_m t} + e^{i\varphi_3}e^{-i\omega_m t})e^{-i\omega t} + (cos\,\epsilon - sin\,\epsilon)\hat{\sigma}_- (e^{i\varphi_1} + e^{i\varphi_2}e^{i\Omega_m t} + e^{i\varphi_3}e^{-i\omega_m t})e^{-i\omega t}) + c.c \quad (2)$$

where $\varphi_1$, $\varphi_2$, and $\varphi_3$ are the phases of the field at frequency $\omega_0$, $\omega_0 - \omega_m$, and $\omega_0 + \omega_m$ respectively. These can have values 0 or $\pi$ depending on + or –ve phase as discussed earlier. The master equation after applying rotation wave approximation yields time evolution of the density matrix elements as given below

$$\dot{\rho}_{gg} = i(cos\,\epsilon + g\,sin\,\epsilon)\Omega_{ge}\tilde{\rho}_{eg} - i(cos\,\epsilon - g\,sin\,\epsilon)\Omega_{eg}\tilde{\rho}_{ge} + \frac{\Gamma}{3}\rho_{ee} + \sum_{g'\neq g}\gamma\rho_{g'g'} - 2\gamma\rho_{gg} \quad (3)$$

$$\dot{\tilde{\rho}}_{ge} = -i(cos\,\epsilon - g\,sin\,\epsilon)\Omega_{ge}\rho_{gg} + i(cos\,\epsilon - g\,sin\,\epsilon)\Omega_{ge}\rho_{ee} + \sum_{g'\neq g}i(cos\,\epsilon - g'\,sin\,\epsilon)\Omega_{g'e}\rho_{gg'} - gi\Omega_L\tilde{\rho}_{ge} - \frac{\Gamma}{2}\tilde{\rho}_{ge} - i\delta\tilde{\rho}_{ge} - \gamma\tilde{\rho}_{ge} \quad (4)$$

$$\dot{\rho}_{gg'} = i(cos\,\epsilon - g\,sin\,\epsilon)\Omega_{ge}\tilde{\rho}_{eg'} - i(cos\,\epsilon - g'\,sin\,\epsilon)\Omega_{g'e}\tilde{\rho}_{ge} - (g - g')i\Omega_L\rho_{gg'} - 2\gamma\rho_{gg'} \quad (5)$$

$$\dot{\rho}_{ee} = -i(cos\,\epsilon - g\,sin\,\epsilon)\Omega_{ge}\tilde{\rho}_{eg} + i(cos\,\epsilon - g\,sin\,\epsilon)\Omega_{eg}\tilde{\rho}_{ge} - \Gamma\rho_{ee} \quad (6)$$

where $\Omega_{ge} = e^{i\varphi_1}\Omega_0^{ge} + e^{i\varphi_2}\Omega_{-1}^{ge}e^{i\omega_m t} + e^{i\varphi_3}\Omega_{+1}^{ge}e^{-i\omega_m t}$ is the total Rabi frequency. $\Omega_0^{ge} = \frac{\mu_{ge}.E_0}{\hbar}$, $\Omega_{-1}^{ge} = \frac{\mu_{ge}.E_{-1}}{\hbar}$, and $\Omega_{+1}^{ge} = \frac{\mu_{ge}.E_{+1}}{\hbar}$ are the Rabi frequency of the component at $\omega_0$, $\omega_0 - \omega_m$, and $\omega_0 + \omega_m$ respectively. The $E_0$, $E_{-1}$, and $E_{+1}$ are their respective electric field. The $e$ and $g$ (or $g'$, with $g \neq g'$) represents the excited and ground level Zeeman sub-states (Fig.2B). $\tilde{\rho}_{ge} = \dot{\rho}_{ge}e^{i\omega t}$, $\delta$ is frequency detuning, and other symbol have their customary meaning. To solve the above-coupled equations for steady-state, we expand the density matrix elements $\rho_{jk}$ in a Fourier series as

$$\rho_{jk} = \sum_{n=-\infty}^{\infty}\rho_{jk}^{(n)}e^{in\omega_m t} \quad (7)$$

where j, k are ground and excited states, $\rho_{jk}^{(n)}$ are slowly varying amplitudes. The Eq. 7 is substituted into Eq. 3-6 with a condition $\rho_{ee} + \sum\rho_{gg} = 1$. A resultant set of 15 recurrence equations are obtained as $n$ varies from $-\infty$ to $+\infty$. They get reduce to 9 equations as $m_J = 0$ is not coupled to the excited state. The set of 9 coupled recurrence equations are solved by using method described in the Ref-25. The slowly varying amplitudes $\rho_{jk}^{(n)}$ are constructed in the form of a column vector $\boldsymbol{\rho}$ having a dimension $9 \times (2N + 1)$ as

$\left(\rho_{gg}^{(-N)},\rho_{gg'}^{(-N)},\rho_{g_ng_n}^{(-N)},\rho_{ge}^{(-N)},\rho_{eg}^{(-N)},\ldots\ldots,\rho_{gg}^{(0)},\rho_{gg'}^{(0)},\rho_{g_ng_n}^{(0)},\rho_{ge}^{(0)},\rho_{eg}^{(0)},\ldots\ldots,\rho_{gg}^{(N)},\rho_{gg'}^{(N)},\rho_{g_ng_n}^{(N)},\rho_{ge}^{(N)},\rho_{eg}^{(N)}\right)^T$,

where $g, g'$ are the ground Zeeman sub-state but $g \neq g'$, $g_n$ is the non-coupled ground Zeeman sub-state and $e$ is the excited state. We restrict $N = 2$ as higher order doesn't contribute for a trichromatic field. The set of equations for slowly varying amplitudes under steady state conditions can be written as

$$\boldsymbol{Q\rho} = \boldsymbol{R} \quad (8)$$

where $\boldsymbol{Q}$ is a $9 \times (2N + 1)$ by $9 \times (2N + 1)$ matrix and $\boldsymbol{R}$ is $9 \times (2N + 1)$ column vector. We have used the matrix inversion method to get the steady state solution for $\boldsymbol{\rho}$.

The atomic polarization in the spherical coordinate system is given by $P \propto Tr(\rho d)$. The absorption, birefringence, and dichroism at steady-state are calculated from the following expressions:

Absorption $\propto \frac{Im[\rho_{-1,0'}]}{(cos\,\epsilon + sin\,\epsilon)\Omega_{-10'}} + \frac{Im[\rho_{1,0'}]}{(sin\,\epsilon - cos\,\epsilon)\Omega_{10'}}$ (9)

Birefringence $\propto \frac{Re[\rho_{-1,0'}]}{(cos\,\epsilon + sin\,\epsilon)\Omega_{-10'}} - \frac{Re[\rho_{1,0'}]}{(cos\,\epsilon - sin\,\epsilon)\Omega_{10'}}$ (10)

Dichroism $\propto \frac{Im[\rho_{-1,0'}]}{(cos\,\epsilon + sin\,\epsilon)\Omega_{-10'}} - \frac{Im[\rho_{1,0'}]}{(sin\,\epsilon - cos\,\epsilon)\Omega_{10'}}$ (11)

The experimentally observed signal profiles are the negative first derivative of the actual line shape profile due to the use of a phase-sensitive detection technique (with a phase shift of $180^0$). It is accommodated by displaying the negative derivative of the calculated line shape profile in this article. The calculation uses $\Gamma = 5.6$ MHz, $\gamma = 1$kHz, and $\Omega_0^{ge} = \Omega_{-1}^{ge} = \Omega_{+1}^{ge} = 5$kHz. The smaller value of the Rabi frequency is due to distribution of the laser power among the teeth component, but here it has only phenomenological implication. The signal profile deviates from the experimental line shape for higher value of Rabi frequency (> 100 kHz).

The calculated absorption using Eq.9 shows different behaviour for the two kinds of phase relationships in Fig. 3. The sequence of relationships that exists at the central part of the comb (Fig. 2B) does not show the magnetic resonance at $\Omega_L = \pm\,\omega_m/2$. This behaviour is consistent with the prior experimental observations [19]. All the resonances ($\Omega_L = 0, \pm\,\omega_m/2$, and $\pm\omega_m$) are observed for the other set of phase relationships. The qualitative agreement between the calculation and the measurement for $\Delta\sim+140$ is shown in the figure. The experimental conditions for the measured profile in Fig.3 are maintained in the rest of this work. The resemblance is an improvement from the erstwhile model (using full spectrum of FM light field), where the inconsistency was circumvented by carrying out the calculations with a smaller modulation index as compared to the experimental value [19]. The trichromatic field model overcomes this inconsistency and indicates that the quantum interference between the first and second neighbour teeth components contributes to the signal prominently for the utilized experimental parameters. Thus, the presented model is not only simple but also appropriate for the employed experimental parameters. However, the measured profile depends on the modulation parameters. The magnetic resonances at higher harmonics of $\pm\omega_m$ appear on further increasing the $A_m$ from the utilized value. The trichromatic field will have obvious limitations to explain these coherences as they are due to the coupling between distant teeth components. Nevertheless, the simplified model is better suited to explains the fundamental aspect of the phenomena.

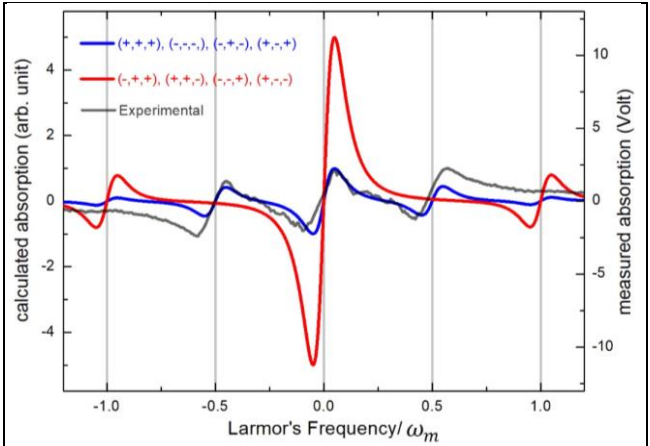

**Fig. 3:** The consistency of the theoretical model with the experimental observation is presented. The calculation is carried out for the possible phase relationship between the three fields. The distinct feature of the two groups is consistently captured in the calculated profile (blue and red curve) for the magnetic component at $\Omega_L = \pm\omega_m/2$. The measured change in absorption (MMz) profile for $\Delta\sim+140$ MHz from the $^{85}$Rb $F = 3 \to F' = 2$ transition is shown by the grey curve. The calculated profile by the trichromatic field (blue curve) is qualitatively consistent with the measured data.

The above discussion is related to the $\Delta$ dependence of the signal profile. The relative phases among the trichromatic field are changed in the model to address the interaction of the atomic ensemble with the different parts of the NBFC. Experimentally, the detuning of the central component of NBFC is represented by $\Delta$. The trichromatic field itself can interact resonantly or off-resonantly with the atom. To avoid confusion with the earlier defined $\Delta$, we present $\delta$ as the detuning of the central component of the trichromatic field from the atomic resonance in the field-free condition. The various attributes of optical activity strongly depend on $\delta = 0$ and $\delta \neq 0$. These are termed as resonant and off-resonant behaviour respectively. The calculated absorption, circular birefringence, and circular dichroism show characteristic line shape depending on the $\delta$ and the ellipticity. Generally, the atomic response is a convolution of all these processes but some of it dominates for specific experimental conditions. The characteristic line shapes contain important information on the underlying physical mechanisms and are summarized in this section. It provides a framework to analyse the experimental observation subsequently. The absorption shows a dispersive line shape at $\Omega_L = 0, \pm\omega_m/2$, and $\pm\omega_m$ in Fig. 4A. The resonant and off-resonant behaviour is identical except for the amplitude. The absorption behaviour is also immune to the change in the ellipticity of the light field.

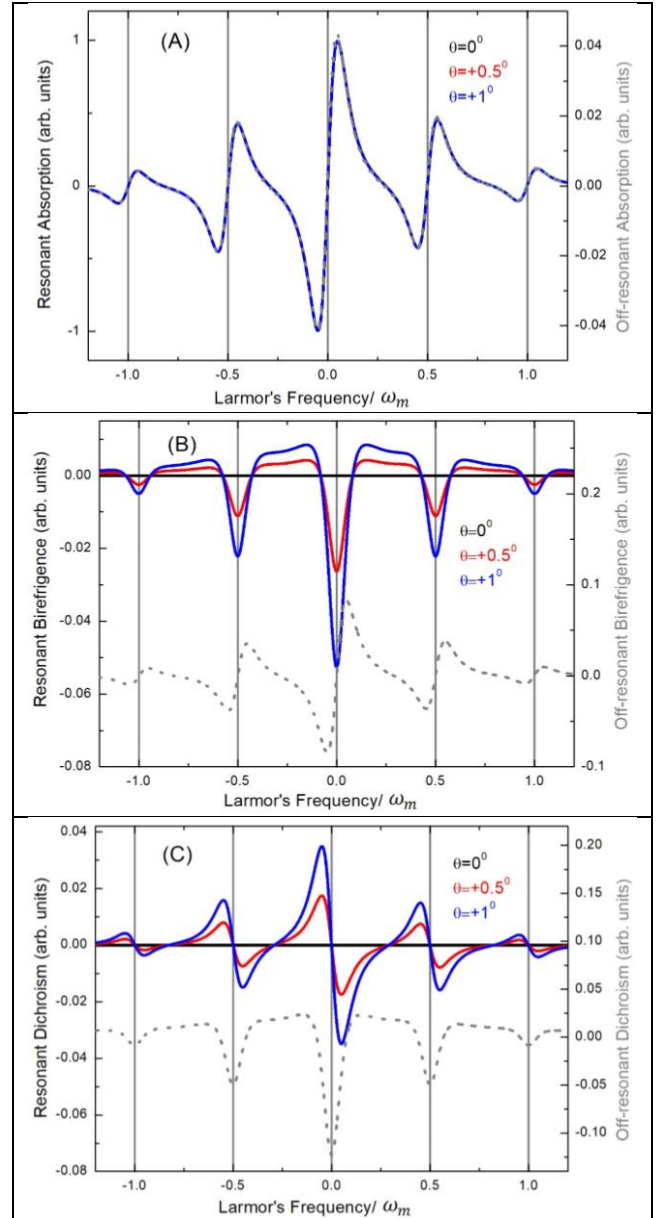

**Fig. 4:** The calculated profile of absorption (A), birefringence (B), and dichroism (C) for different ellipticity of the light field denoted by the angle of the QWP ($\theta$). The solid and dotted line represents resonant ($\delta = 0$) and off-resonant ($\delta = +\Gamma$) behaviour respectively.

The birefringence part of the optical activity is calculated using Eq.10. The circular birefringence for resonant and off-resonant cases have distinct features. There is no circular birefringence for the resonant linearly polarized light field in Fig. 4B. However, the response is very-sensitive at $\Omega_L = 0, \pm\omega_m/2$, and $\pm\omega_m$ for any small change in the ellipticity. The amplitude of the calculated dip structures increases with the ellipticity. It is due to the combined action of the optical pumping and the Zeeman coherence. These dip structures convert to peak structures for the opposite value of θ (not shown here). In relation to the experiment, θ represents the angle between the QWP and PBS-i. In contrast, dispersive line shape at $\Omega_L = 0, \pm \omega_m/2$, and $\pm \omega_m$ are observed for the off-resonant light field. This behaviour is insensitive to the change in the amplitude and polarity of the ellipticity.

The resonant and off-resonant dichroism for different ellipticity of the light field is calculated using Eq.11. There is no resonant dichroism for pure linearly polarized light in Fig.4C. It is due to the inherent symmetry of the system and similar to the observation for resonant birefringence. The dispersive line shape (for $\delta = 0$) at $\Omega_L = 0, \pm \omega_m/2$, and $\pm \omega_m$ increases with the ellipticity. The slope of these line shape gets reversed for the opposite value of θ (not shown here). The off-resonant dichroism has dip structures and insensitive to change in the amplitude and polarity of the ellipticity. Thus, the resonant behaviour for both birefringence and dichroism shows a strong dependency on the ellipticity as compared to the immune response of their off-resonant behaviour. The amplitude of both resonant and off-resonant absorption behaviour show immunity to the change in the ellipticity.

## Experimental results and discussions:

The transmitted and reflected light intensity across the PBS-d in Fig.1 are detected by the PD-t and PD-r respectively. These direct signals (without phase-sensitive detection) for linearly polarized light are shown in Fig.5. The transmitted signal exhibits enhanced transmission as a function of $\Omega_L$. The signal profile contains three partially resolved resonances located at $\Omega_L = 0$ and $\pm \omega_m/2$. The polarization rotation signal in the reflected port is indistinguishable from the ambient noise. The DC offset is due to the residual p-polarized light in the reflection port.

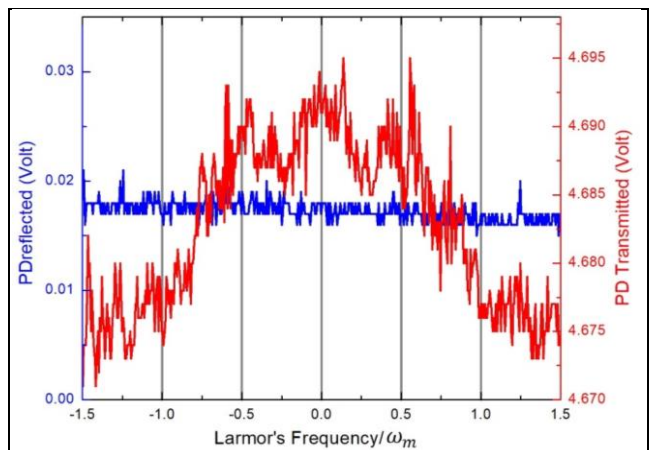

**Fig. 5:** The transmitted (red) and reflected (blue) signal profile (without utilizing phase sensitive detection) across the detection PBS.

The MMzT and MMzR signal in Fig.6 corresponds to the phase sensitively detected transmitted and reflected signal respectively. The signal to noise ratio (SNR) is significantly improved as compared to the direct photo-diode signal in Fig.5. Consequently, the finer-components in both MMzT and MMzR signals are prominently observed. The MMzT signal shows dispersive line shape profile at $\Omega_L = 0$, $\pm \omega_m/2$, and $\pm \omega_m$ as the experiment is carried out at Δ~+140 MHz from the $^{85}$Rb $F = 3 \rightarrow F' = 2$ transition. It corresponds to interaction of the atomic sample with the wing of the NBFC. The measured profile has similar line-shape to the calculated absorption in Fig.4A. The MMzR signal shows three dispersive profile for linearly polarized light at $\Omega_L = 0$ and $\pm \omega_m/2$. The components at $\pm \omega_m$ and beyond it are not resolved with the current experimental parameters. The MMzR signal profile is extremely sensitive to the alignment of the PBS-d and purity of the light polarization. The system is symmetric for a resonant, linearly polarized tri-chromatic field. Thus, there is no circular birefringence or dichroism at resonance as in Fig.4B and C. This allows us to observe the feeble off-resonant behaviour prominently. The MMzT and MMzR signal have similar dispersive profile as well as same polarity $\Omega_L = 0$ and $\pm \omega_m/2$. The off-resonant birefringence has similar agreement with the calculated absorption profile as in Fig.4. Thus, the polarization rotation observed for linearly polarized light is due to enhanced birefringence (off-resonant) arising due to the establishment of Raman resonance at multiple $\Omega_L$.

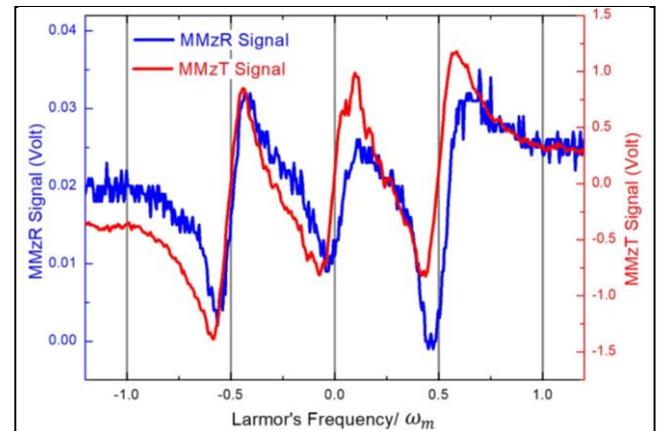

**Fig. 6:** The experimentally observed MMzR and MMzT signal for linearly polarized light field (θ~0⁰). The experiment is carried out at Δ~+140 MHz from the $^{85}$Rb $F = 3 \rightarrow F' = 2$ transition.

The ellipticity of the light field provides a handle for controlling the extent of quantum interference assisted absorption, birefringence, dichroism, and optical pumping. There is a contribution from zero-field Zeeman redistribution also, but it is not incorporated in the model. So, it is discussed as an additional mechanism while interpreting the experimental results. For elliptically polarized light, the optical pumping drives population imbalances between the relevant ground Zeeman levels that further augment the optical activity. There is a technical issue associated with the experimental configuration while changing the ellipticity of the light field. The rotation of QWP by θ changes the transmitted and reflected light intensity across PBS-d proportional to $[1 + \cos^2 2\theta]$ and $\sin^2 2\theta$ respectively. Thus, the assignment of absorption and polarization rotation to the MMzT and MMzR signal respectively is compromised. For small values of θ, this contribution is minuscule and is not a part of the discussion.

For small values of θ (~±1⁰), the MMzT signal does not show any noticeable changes in Fig. 7. The calculated absorption profile using Eq.9 shows consistent behaviour in Fig. 4A. However, the MMzR signal undergoes remarkable enhancement even for small variation in θ. Also, the MMzR signal has improved SNR than the MMzT signal due to reduced intensity noise in the MMzR signal. The distinction in SNR is substantial for θ ~ ±3⁰ (not shown here). Thus, the MMzR signal has better prospects for metrological application.

The MMzR signal shows dispersive line shape at $\Omega_L = \pm \omega_m/2$ for linearly polarized light. It changes to a peak or dip structure for θ ~ +1⁰ or -1⁰ respectively as in Fig.7. The nature of the line shape is similar to the resonant birefringence in Fig.4B. The increase in the signal amplitude is due to the population imbalance, driven by the optical pumping. The measured change in the polarity of the signal for the opposite θ is consistent with the calculation (not shown here). It indicates the reversal of the relative phase shift between the left and right circularly polarized light fields for opposite value of θ. Thus, resonant birefringence suitably explains the behaviour of the MMzR signal at $\Omega_L = \pm \omega_m/2$ for small value of θ. The off-resonant birefringence is obscured (the small asymmetry in the line shape may be noticed) due to the strong resonant birefringence under the prevailing circumstances.

The polarity of the MMzR signal at $\Omega_L = 0$ in Fig.7 is opposite to the calculated resonant birefringence in Fig. 4B. The calculated off-resonant dichroism in Fig. 4C has similar peak line shape as the above measurement but is immune to the change in the ellipticity. Thus, it cannot be the underlying mechanism as the MMzR signal at $\Omega_L = 0$ has strong dependency on the ellipticity of the light field. The explanation of the line shape at $\Omega_L = 0$ requires the consideration of simultaneous action of off-resonant birefringence, resonant birefringence, and Zeeman redistribution. The last process is a peculiarity of zero field ($\Omega_L = 0$) phenomena and is associated with enhanced absorption. It will correspond to a dispersive line shape profile with a negative slope at the line centre. In this parametric regime, the Zeeman redistribution drives the mechanism through a different route. It destroys the asymmetry in the ground state populations generated by the optical pumping near $\Omega_L=0$. Consequently, the optical activity of the medium gets compromised, and there is a reduction in the MMzR signal at $\Omega_L = 0$. The apparent peak structure is a convolution of the above processes. However, the complicacy of the Zeeman redistribution is absent for $\Omega_L = \pm \omega_m/2$ and the MMzR signal is consistent with the calculated on-resonance birefringence. The on-resonant birefringence is also responsible for the magnetic field shift of the magnetic resonances at $\Omega_L=0$ and $+\omega_m/2$ in Fig. 6. It is due to a feeble ellipticity that gets incur in the light field during the scanning of the magnetic field.

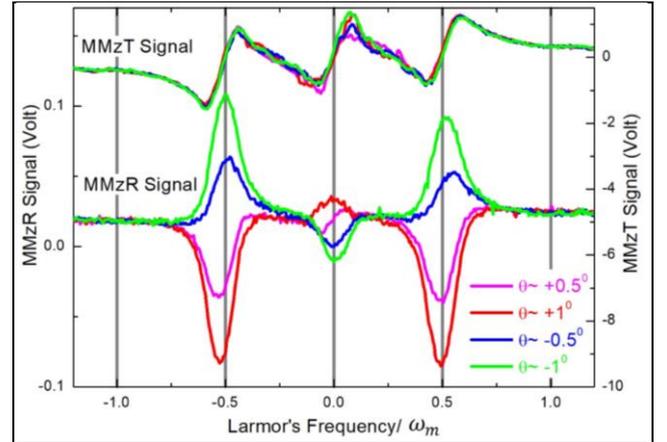

**Fig. 7:** The response of the MMzR and MMzT signal profiles to a small change in the ellipticity of the input light field. The MMzT signal shows no difference whereas MMzR signal is extreme sensitive to a small change in ellipticity.

For a small value θ, the MMzR signal closely corresponds to the polarization rotation signal. However, some part of the absorption gets mixed with the MMzR signal at higher value of θ. Thus, the MMzR signal is a convolution of change in the light intensity (proportional to $\sin^2 2\theta$ as discussed earlier) and change in light polarization, for higher value of ellipticity. Similarly, the modification in the polarization of the light field also contributes to the MMzT signal profile. The MMzR, MMzT, and their combined signal for θ=±20⁰ and ±45⁰ are shown in Fig. 8 A, B and C respectively.

The dispersive profile at $\Omega_L=0$ in the MMzR signal for θ=±20⁰ is due to the optical pumping followed by Zeeman redistribution (Fig. 8A). The negative slope at the line centre represents enhanced absorption and is verified by consistent behaviour in the direct photodiode signal. Since the Zeeman redistribution annuls the optical pumping induced asymmetry, the optical activity of the medium gets compromised. Thus, there is limited role of polarization rotation in the line shape at $\Omega_L=0$ in the present circumstances. The resonance structure at $\pm\omega_m/2$ for θ=±20⁰ is primarily due to resonant birefringence augmented by enhanced optical pumping. Consequently, the optical rotation has nearly an order of higher amplitude than the corresponding signal in Fig. 7. The asymmetric peak structure and its small shift from the expected $\pm\omega_m/2$ are due to the addition of the absorption part (dispersive line shape) in the signal as discussed in the previous paragraph.

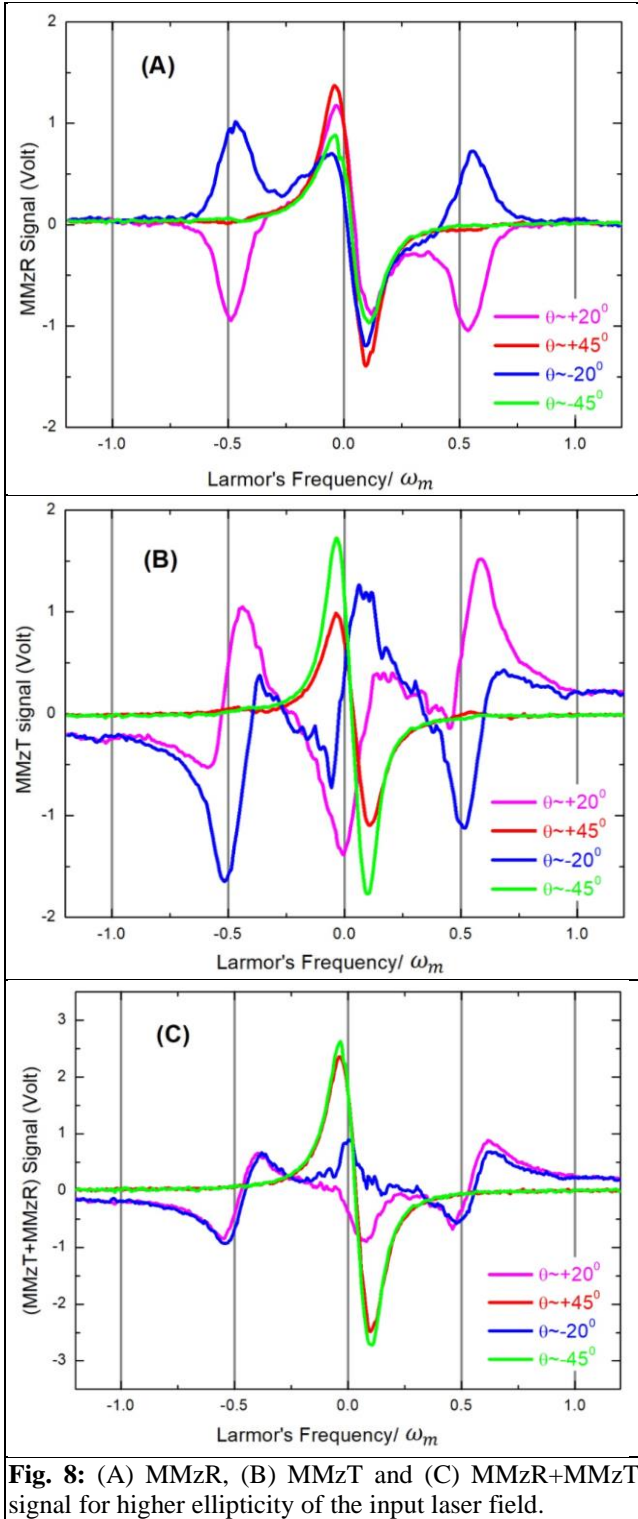

**Fig. 8:** (A) MMzR, (B) MMzT and (C) MMzR+MMzT signal for higher ellipticity of the input laser field.

The resonance structure at $\Omega_L=0$ in the MMzT signal for $\theta=\pm20^0$ (Fig.8B) is a convolution of change in absorption due to quantum interference and Zeeman redistribution of the ground state population. The first mechanism leads to enhanced transmission, whereas the second one is responsible for enhanced absorption. The asymmetric dispersive profiles are observed at $\pm\omega_m/2$ in the MMzT signal. It is a convolution of change in absorption due to quantum interference and reduction in light intensity in the transmission port of PBS-d due to polarization rotation.

Fig. 8C shows the combined MMzR and MMzT signals for $\theta=\pm20^0$ and $\pm45^0$. The PBS is used to discriminate the change in polarization of the light field. The combined signal represents the change in light intensity only. It retrieves the symmetry of the dispersive profile at $\Omega_L = \pm\omega_m/2$ resonance structure for $\theta=\pm20^0$. The apparent peak function at $\Omega_L = 0$ (for $\theta=\pm20^0$) is a convolution of two dispersive profiles with opposite slope, one due to quantum interference assisted enhanced transmission and the other due to Zeeman redistribution assisted enhanced transmission. There is a small separation in the line centre of these two mechanisms. Consequently, the convolution of resonance structures gives rise to an apparent peak profile. The observation infers nearly equal amplitude of these two mechanisms for $\theta=\pm20^0$.

The MMzT and MMzR signal for $\theta=\pm45^0$ has a reciprocal relationship in amplitude as in Fig.8A and B. The combined signal has a similar magnitude for both polarization as in Fig. 8C. There is no quantum interference for a pure circularly polarized light field, and Zeeman redistribution is the only possible mechanism in absence of the orthogonal magnetic field. Thus, a single dispersive line shape is observed for $\theta=\pm45^0$. As a small orthogonal field is applied, magnetic resonances start appearing at $\Omega_L = \pm\omega_m/2$ in both MMzR and MMzT signal (not shown here), which vanishes in the combined signal. Thus, the absence of resonances at $\Omega_L = \pm\omega_m/2$ (in the MMzR and MMzT signal for $\theta=\pm45^0$) supports proper compensation of the orthogonal field in the presented work. The observation also indicates the existence of Zeeman redistribution even in absence of an orthogonal magnetic field.

## Conclusions:

The complex interaction of NBFC with an atomic system is suitably addressed by a trichromatic field model. It provides a microscopic picture of the associated physical processes. The model is validated by the fundamental aspects of the phenomena, like the position of the magnetic resonances and their dependencies on the relative phase between the field components. The success of the model indicates the dominating role of the atomic coherence established by the neighbouring field component of the NBFC behind the phenomena for the utilized experimental parameters. The enhanced optical activity (tailored by the NBFC) near multi-photon resonances are measured. It provides a rich system to study the fundamental aspect of optical activity in a variety of conditions. The role of birefringence, dichroism, quantum interference assisted enhanced transmission, and Zeeman redistributing assisted population redistribution in the observed signal profile is discussed. The simple trichromatic field model infers the possible existence of similar steady-state atomic polarization while using an amplitude-modulated light field.

**Acknowledgements:**

The authors are thankful to Dr. Archana Sharma, AD, BTDG for supporting the activity.